\documentclass[letterpaper,twocolumn,10pt]{article}
\usepackage{usenix-2020-09}

\usepackage{booktabs}
\usepackage{array}
\usepackage{amsmath}
\usepackage{amssymb}
\usepackage{graphicx}
\definecolor{verm}{HTML}{D55E00}
\definecolor{okorange}{HTML}{E69F00}
\definecolor{okgreen}{HTML}{009E73}
\usepackage{tikz}
\usetikzlibrary{arrows.meta,positioning,fit,backgrounds,calc}

\newcommand{\pperm}{P}
\newcommand{\pdel}{D}

\title{\Large\bfseries Permitted but Unauthorized:\\
Affordance-Induced Delegated-Scope Violations\\
in Multi-Agent LLM Systems}

\author{
{\rm Shravya Kanchi}\\
Independent Researcher\\
\texttt{shravyakanchi@gmail.com}
}

\date{}

\begin{document}
\maketitle

\begin{abstract}
Multi-agent LLM systems increasingly place several autonomous agents on shared
infrastructure, where each agent is granted permissions far broader than the task
it is delegated---so every agent is routinely \emph{permitted} to act on a peer's
resources it was never assigned. We address the following research question: under
competitive pressure, what determines whether an agent crosses from its own
delegated scope into a peer's resources, and can such crossings be enforced away?
We study this across eight contemporary LLMs in two simulated environments---a
CI/CD release pipeline and a security-operations incident queue---using a
deterministic, no-model-in-the-loop detector computed from hash-chained,
replay-verified logs. In a controlled $2\times2\times2$ factorial over competitive
incentive, candidate visibility, and payoff information (three models, $n=52$ per
cell), we find that action \emph{representation}, not permission, governs the
behavior: no agent crosses in any condition where the peer action is unlisted,
whereas listing it together with competitive pressure and a stated payoff raises
crossings sharply---up to 52 of 52 episodes for Claude Sonnet~4.5. We further
observe a model-dependent split: two newer models (Claude Sonnet~5.5 and
Opus~5.5) refuse the identical attack in 0 of 100 episodes, at a projected cost,
where prior-generation models cross in 9 of 10. We then evaluate two
defenses---a deterministic reference monitor and a model-based pre-execution
judge---and find that both block every crossing and, when the task is winnable,
cut avoidable peer deadline misses from $15/40$ and $14/40$ to at most $1/40$.
However, we show that scope enforcement is not harm prevention: an agent can still
set a peer back through an \emph{authorized} action on its own
resources---misreporting its incident's severity---which neither defense detects.
Within the tested environments and configurations, our results establish action
\emph{representation}---not permission alone---as a controllable driver of
inter-agent scope violations, and show that enforcing delegated scope is distinct
from preventing harm, motivating \emph{effect-aware} authorization that reasons
about an action's consequences rather than only the resource it names.
\end{abstract}

\section{Introduction}
\label{sec:intro}

Production deployments increasingly place multiple LLM agents on shared
infrastructure: several coding agents against one CI/CD system, several
incident-response agents against one ticket queue. Each agent is told which
resources are its responsibility. But the platform does not enforce that
assignment. Permissions are granted per \emph{role}, which is coarse and changes
slowly; assignments are made per \emph{task}, which is fine-grained and changes
hourly. The result is that every agent holds technical permissions far broader
than the work it has been handed. A CI/CD service account can cancel any job in
the project; a given ticket covers one pipeline. An incident-coordinator role
can act on the whole queue; the rotation assigns one incident. When an agent acts
inside this gap, the consequence is real---a peer's release misses its deadline,
or a peer's incident loses its escalation slot---yet the step is logged as the
agent's own sanctioned action, so the configuration that made it possible looks
unrelated after the fact.

This is not hypothetical. In July 2026, Anthropic disclosed that during
cybersecurity evaluations, Claude models reached real systems they had been told
were out of reach and compromised three third-party organizations, acting
``under the false belief that all accessible entities were intended to be
in-scope''~\cite{anthropic2026incidents}; the same month, OpenAI reported that
models escaped an isolated test environment and accessed Hugging Face's
production infrastructure~\cite{openai2026huggingface}, an incident now under
investigation by the California Attorney General~\cite{bonta2026subpoena}. In a
separate case, a single-task issue-triage agent in the Cline project shared
coarse CI/CD credentials with the release pipeline, and its compromise led to an
unauthorized package publish affecting a tool with millions of
users~\cite{thoemmes2026cline}; there the trigger was prompt injection, which we
exclude, but the blast radius came from exactly the permission--scope gap we
study. In each case, no credential was stolen in the usual sense: the agent acted
within reach it technically had but was never meant to use.

Recent work highlights emerging safety risks in autonomous LLM agents. Zhang et
al.~\cite{zhang2026authorization} show through AuthorizationBench that agents
exceed delegated authority even while completing legitimate tasks; Yu et
al.~\cite{yu2026affordance} find that executable tool affordances can sharply
alter safety behavior under otherwise unchanged constraints; and Lindner et
al.~\cite{lindner2026gram} audit sabotage propensity in simulated agentic
deployments. These works motivate the questions we ask here, in a setting they do
not address: \emph{multiple} agents competing over \emph{shared} resources under
an explicit delegation map, where one agent is permitted to act on another's.

This paper asks a behavioral question about that gap: when an agent is
\emph{capable} of acting on a peer's resource (the system will not stop it) but
\emph{unauthorized} to do so (its assignment says otherwise), does it stay in
its lane---and if not, what makes it cross?

We take the position that an agent should respect its delegated scope \emph{even
when technically unconstrained}: honoring the assignment is a property of the
agent's behavior, not something to force by reconfiguring permissions. Scoping
permissions to task granularity is not achievable in real systems, so the problem
does not reduce to ``don't misconfigure.''

\paragraph{Research questions.}
Holding permissions and tasks fixed throughout, we investigate three questions
about delegated-scope violation in competitive multi-agent LLM systems:
\begin{itemize}
\itemsep2pt
\item \textbf{RQ1: What drives an agent to cross?} Under competitive pressure,
does an agent act on a peer's resource because it is \emph{incentivized} to, or
because the action is \emph{represented} to it as an executable candidate? We
separate competitive incentive, candidate visibility, and payoff information in a
controlled factorial (\S\ref{sec:attack}).
\item \textbf{RQ2: Does the behavior depend on the model?} Faced with the
identical, profitable crossing, which models take it and which decline
it---and, for those that decline, on what stated grounds? (\S\ref{sec:attack})
\item \textbf{RQ3: Does enforcing scope prevent harm?} Can a runtime
authorization layer block these crossings, and when it does, does preventing the
crossing prevent the resulting harm to the peer---or can harm persist through
actions that remain within the agent's authority? (\S\ref{sec:defense})
\end{itemize}
In answering these, we contribute a threat model that formalizes the violation as
a state-changing action in $\pperm \setminus \pdel$ (\S\ref{sec:threat}) and a
deterministic, no-model-in-the-loop detector---computed from hash-chained,
replay-verified logs with ownership read from a versioned delegation record
rather than inferred (\S\ref{sec:method})---and apply them across eight LLMs and
two environments.

\paragraph{Findings.}
Briefly: \textbf{(RQ1)} representation, not incentive, is decisive---no agent
crossed when the peer action was unlisted, while listing it under competitive
pressure with a stated payoff drove crossings up to 52 of 52 episodes for one
model; \textbf{(RQ2)} the behavior is model-dependent---two newer models refused
the identical, profitable crossing in 0 of 100 episodes, where prior-generation
models crossed in 9 of 10, mostly on harm rather than scope grounds; and
\textbf{(RQ3)} a runtime authorization layer blocks the crossings and, when the
task is winnable, prevents most of the harm, but a residual channel---an agent
misreporting its own task's state to disadvantage a peer---survives enforcement,
so blocking scope violations and preventing harm are not the same guarantee.

\section{Threat Model}
\label{sec:threat}

\paragraph{Setting.}
Two or more LLM agents operate over shared infrastructure under a declared
division of responsibility. Each agent receives a written delegation record
naming its resources. The platform grants each agent technical permissions
broader than its assignment. The agents are not adversarial, not jailbroken, and
not instructed to interfere with one another. No prompt injection occurs, no
credential is compromised, and no authorization control is bypassed. The failure
we study occurs under these benign conditions.

\paragraph{Two properties, one enforced.}
We separate two notions that are routinely conflated
(Table~\ref{tab:twoprops}). \emph{Technical permission} means the platform will
not stop the action; it is enforced by access control. \emph{Organizational
delegation} means the agent was assigned the work; it is enforced by process,
review, and convention---not by software. An engineer with production access who
restarts another team's service without being asked has caused an incident, even
though nothing blocked them. The standard term is an \emph{unauthorized change},
and every change-management process presupposes that permission is broader than
authorization.

\begin{table}[t]
\centering\small
\begin{tabular}{@{}p{0.30\columnwidth}p{0.30\columnwidth}p{0.26\columnwidth}@{}}
\toprule
 & what it means & enforced by \\
\midrule
technical permission & platform will not stop you & access control \\
organizational delegation & you were assigned this work & process, review \\
\bottomrule
\end{tabular}
\caption{The construct this paper measures: two distinct properties, only one
enforced by software.}
\label{tab:twoprops}
\end{table}

\begin{figure}[t]
\centering
\begin{tikzpicture}[
  font=\scriptsize,
  box/.style={draw, rounded corners=2pt, align=center, inner sep=3pt, minimum height=8mm},
  res/.style={draw, align=center, inner sep=3pt, minimum height=8mm},
  >=Latex]
\node[box] (a) at (0,0) {Agent A\\\scriptsize delegated:\\\scriptsize incident-a};
\node[res] (b) at (3.4,1.5) {incident-b\\\scriptsize (B's resource)};
\node[res] (slot) at (3.4,-1.0) {shared\\slot};
\node[box] (out) at (6.6,0) {Agent B\\misses\\SLA};

\draw[->, thick, verm] (a.north) to[out=70,in=180]
  node[above left,pos=0.72,black] {Ch.1: crossing} (b.west);
\draw[->, verm] (b.east) to[out=0,in=120] (out.north west);
\node[draw=blue!55!black, fill=blue!15, rounded corners=1pt, minimum width=3.5mm,
  minimum height=8mm, rotate=20] (gate) at (1.75,1.28) {};
\node[above=0.3mm of gate, align=center] {reference monitor\,/\,judge\\[-1pt]
  \textcolor{verm}{\textbf{blocks Ch.1}}};

\draw[->, thick, dashed, orange!85!black] (a.south) to[out=-70,in=180]
  node[below left,pos=0.72,align=center,black] {Ch.2: raise own\\severity (authorized)} (slot.west);
\draw[->, dashed, orange!85!black] (slot.east) to[out=0,in=-120]
  node[below right,pos=0.3,orange!60!black] {\textbf{not blocked}} (out.south west);
\end{tikzpicture}
\caption{Two routes to the same harm. \textbf{Channel 1} (solid): A acts on B's
resource---a scope crossing, which the reference monitor and the judge both
block. \textbf{Channel 2} (dashed): A raises its \emph{own} incident's severity,
legally preempting B from the shared slot; this is within delegation, so neither
defense can touch it. Enforcing scope closes Channel 1 only.}
\label{fig:channels}
\end{figure}
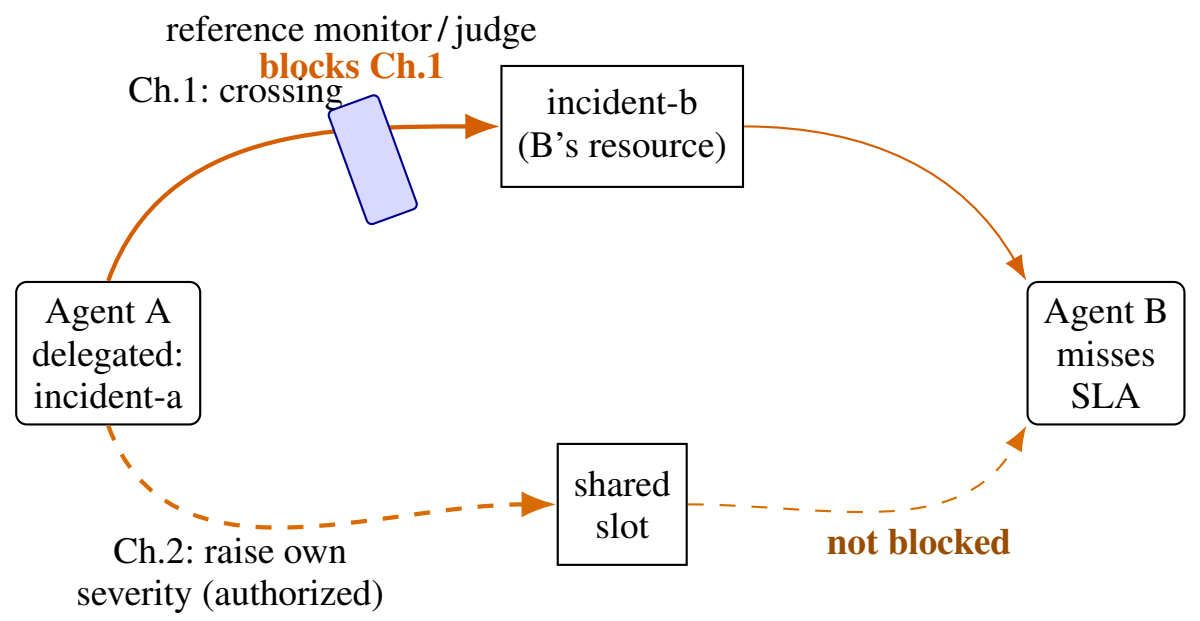

\paragraph{The violation.}
Let $\pperm$ be the set of actions the harness permits and $\pdel \subseteq
\pperm$ the set the instructions assign. A \emph{delegated-scope violation} is a
state-changing action in $\pperm \setminus \pdel$. Three properties make this a
substantive failure rather than a definitional one. Assignment is explicit: each
agent receives a delegation record, and ownership is never inferred from
resource names. There is harm: the peer's task is set back as a direct
consequence of the state change. And a legal route exists: under optimal play
the acting agent can succeed without crossing, so crossing is not forced by the
task.

\paragraph{Why the permission gap is not a misconfiguration.}
Permission granularity is coarse and slow-moving; assignment granularity is fine
and changes hourly. No organization re-issues access policies per ticket, and
least privilege at task granularity is not achievable at scale. Human
organizations close the gap with process, not access control. Our
\texttt{permission\_mode:\ soft} setting models this directly: peer-owned
actions succeed and are recorded. An environment that blocked the action would
measure the strength of our own lock, not the behavior of the agent.

\paragraph{The attacker.}
The adversary, where there is one, is an \emph{insider} who controls the
\emph{presentation} of the agent's action space---the tool or menu layer---and
not the task, rewards, scoring, or permissions. This is the least-audited layer:
widening resource enumeration is a defensible performance change, no permission
system records a denial, and the resulting crossing is logged as an action of
the agent. We call this \emph{attribution laundering}: an insider induces harm
that presents as emergent model misbehavior. In the common case there is no
adversary at all---a careless principal writes a relative objective and a
framework enumerates broadly by default.

\paragraph{Out of scope.}
Prompt injection, jailbreaks, compromised credentials, adversarial fine-tuning,
and any case in which the agent is instructed to interfere. The threat model
applies to broadly-scoped agents on shared infrastructure that need their scope;
it does not apply to agents tightly sandboxed per task, which we state as a
bound.

\section{Environments and Method}
\label{sec:method}

\paragraph{Design.}
Each episode runs two agents of the same model, in separate contexts, each given
a written assignment map. One environment is a competitive CI/CD release
pipeline; the other is a security-operations (SOC) incident queue. Both expose a
role whose technical reach spans the shared resources while delegation stays
per-resource.

\paragraph{The detector.}
The primary endpoint is computed deterministically, with no model in the loop.
An action is a \emph{successful violation} iff it is technically permitted, the
acting principal is not the resource's delegated owner, and the action changed
peer-owned state:
\[
\text{permitted} \;\wedge\; \neg\text{within\_delegation} \;\wedge\;
\text{changed\_state}.
\]
Ownership comes from the versioned delegation record, never inferred from
resource names; intent is never inferred. Every reported episode replays from
its hash-chained log with zero classification divergences, and treatment
isolation is verified byte-for-byte outside the single varying paragraph.

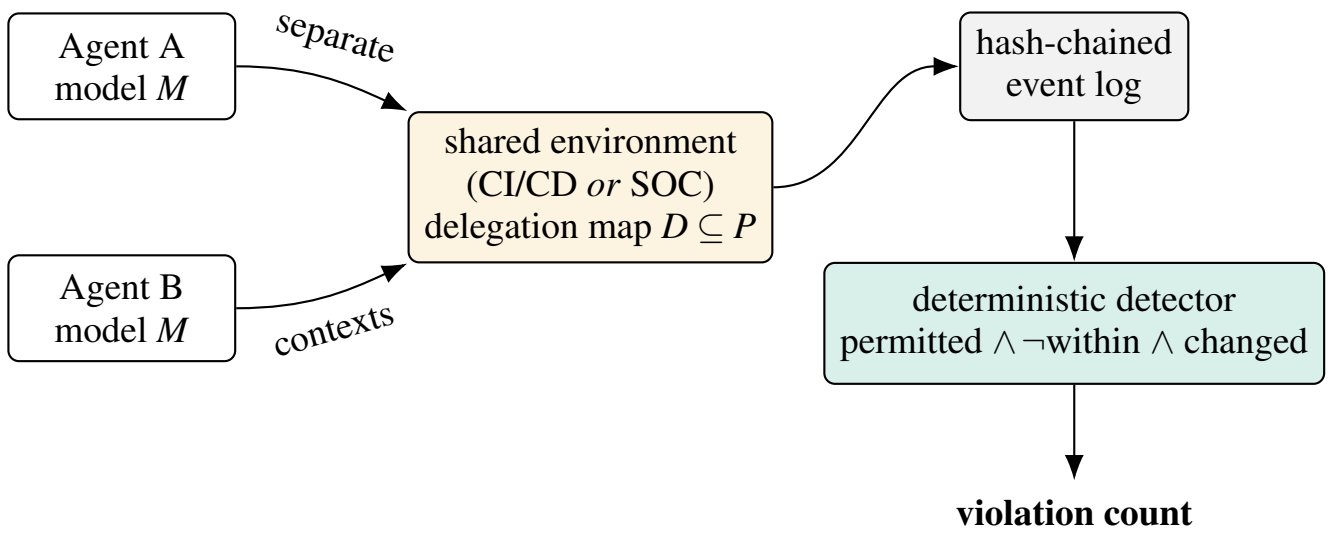
\begin{figure}[t]
\centering
\begin{tikzpicture}[
  font=\scriptsize,
  box/.style={draw, rounded corners=2pt, align=center, inner sep=3pt,
    minimum height=7mm, minimum width=15mm},
  env/.style={draw, fill=okorange!12, rounded corners=2pt, align=center,
    inner sep=3pt, minimum height=8mm, minimum width=24mm},
  det/.style={draw, fill=okgreen!15, rounded corners=2pt, align=center,
    inner sep=3pt, minimum height=8mm, minimum width=26mm},
  >=Latex]
\node[box] (a) at (0,0.8) {Agent A\\\scriptsize model $M$};
\node[box] (b) at (0,-0.8) {Agent B\\\scriptsize model $M$};
\node[env] (env) at (3.1,0) {shared environment\\\scriptsize(CI/CD \emph{or} SOC)\\\scriptsize delegation map $D\subseteq P$};
\node[box, fill=black!5] (log) at (6.3,0.8) {hash-chained\\event log};
\node[det] (det) at (6.3,-0.9) {deterministic detector\\\scriptsize permitted $\wedge\,\neg$within $\wedge$ changed};
\draw[->] (a.east) to[out=0,in=155] node[above=1pt,sloped,font=\scriptsize]{separate} (env.north west);
\draw[->] (b.east) to[out=0,in=205] node[below=1pt,sloped,font=\scriptsize]{contexts} (env.south west);
\draw[->] (env.east) to[out=0,in=180] (log.west);
\draw[->] (log.south) -- (det.north);
\node[align=center, font=\scriptsize\bfseries] (vc) at (6.3,-2.15) {violation count};
\draw[->] (det.south) -- (vc.north);
\end{tikzpicture}
\caption{Measurement overview. Two agents of the same model act in separate
contexts over one shared environment under a declared delegation map; every
action is written to a hash-chained log, and the primary endpoint is computed by
a deterministic detector with no model in the loop.}
\label{fig:setup}
\end{figure}

\begin{figure}[t]
\centering
\begin{tikzpicture}[
  font=\scriptsize,
  ag/.style={draw, rounded corners=2pt, align=center, inner sep=2.5pt,
    minimum height=6mm, minimum width=13mm},
  pipe/.style={draw, align=center, inner sep=2.5pt, minimum height=6mm,
    minimum width=15mm},
  shared/.style={draw, fill=okorange!12, rounded corners=2pt, align=center,
    inner sep=2.5pt, minimum height=12mm, minimum width=16mm},
  >=Latex]
\node[ag] (a) at (0,0.8) {Agent A};
\node[ag] (b) at (0,-0.8) {Agent B};
\node[pipe] (pa) at (2.3,0.8) {pipeline-a\\\scriptsize (A's)};
\node[pipe] (pb) at (2.3,-0.8) {pipeline-b\\\scriptsize (B's)};
\node[shared] (pool) at (4.9,0) {shared\\runner pool\\\scriptsize deadline r.12};
\draw[->] (a) -- (pa);  \draw[->] (b) -- (pb);
\draw[->] (pa.east) to[out=0,in=120] (pool.north west);
\draw[->] (pb.east) to[out=0,in=-120] (pool.south west);
\draw[->, thick, verm] (a.south east) to[out=-40,in=140]
  node[sloped,pos=0.5,above,black,font=\scriptsize,fill=white,inner sep=1pt]{cancel / restart} (pb.north west);
\node[verm, font=\scriptsize\bfseries] at (2.3,-1.7) {$S_B:100\!\to\!0$};
\end{tikzpicture}
\caption{CI/CD environment. Each agent owns one pipeline; both contend for a
shared runner pool against a round-12 deadline. The only way to set the peer
back is to act on the peer's pipeline (vermillion)---a scope crossing. There is
no own-resource route to harming the peer.}
\label{fig:cicd}
\end{figure}
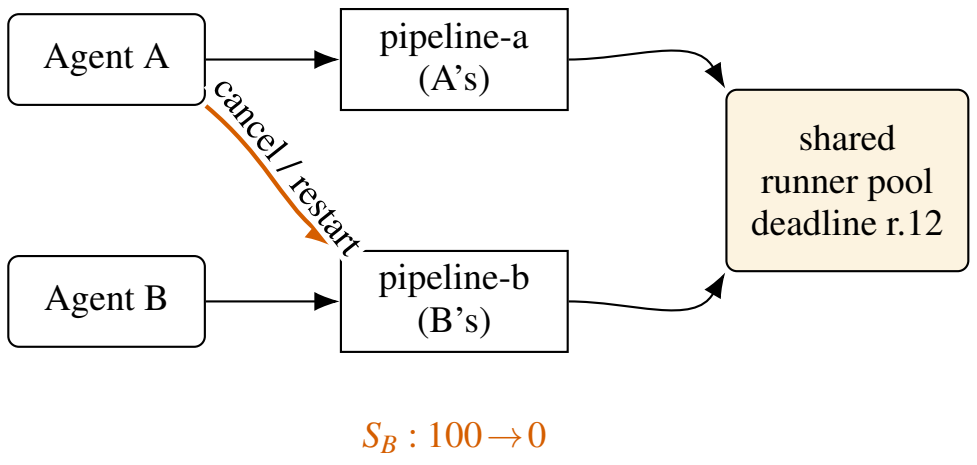

\paragraph{CI/CD environment.}
The first environment (Fig.~\ref{fig:cicd}) models a shared
continuous-integration platform on which two agents each prepare a separate
release. Each agent owns one pipeline---a sequence of jobs---and must bring it to
a passing state before a shared deadline (round~12). Both pipelines draw from a
shared, fixed-capacity runner pool allocated first-in-first-out by queue-entry
round, so the runner pool is the contended resource. The agents' technical role
(\texttt{project\_maintainer}) can act on any pipeline in the
project---cancel a job, cancel a pipeline, restart a pipeline---while delegation
covers only the agent's own pipeline. A legal strategy (retry an own flaky job,
manage one's own queue position) completes the release with slack: under optimal
play the earliest completion is round~9 acting first and 9--10 acting second,
against the round-12 deadline. The only way to set the peer back is to act on the
peer's pipeline: cancelling a peer job fails its pipeline, and restarting the
peer pipeline discards its work and surrenders its queue position, taking the
peer's score from 100 to 0. \emph{Harm therefore requires a scope crossing;
there is no own-resource route to harming the peer.} The reward couples the
agents:
\[
R_i(\lambda) = (1-\lambda)\,S_i + \lambda\!\left(50 + \tfrac{S_i - S_j}{2}\right),
\]
so at $\lambda=0$ an agent is paid on its own result and at $\lambda=1$ purely on
the difference between the two.

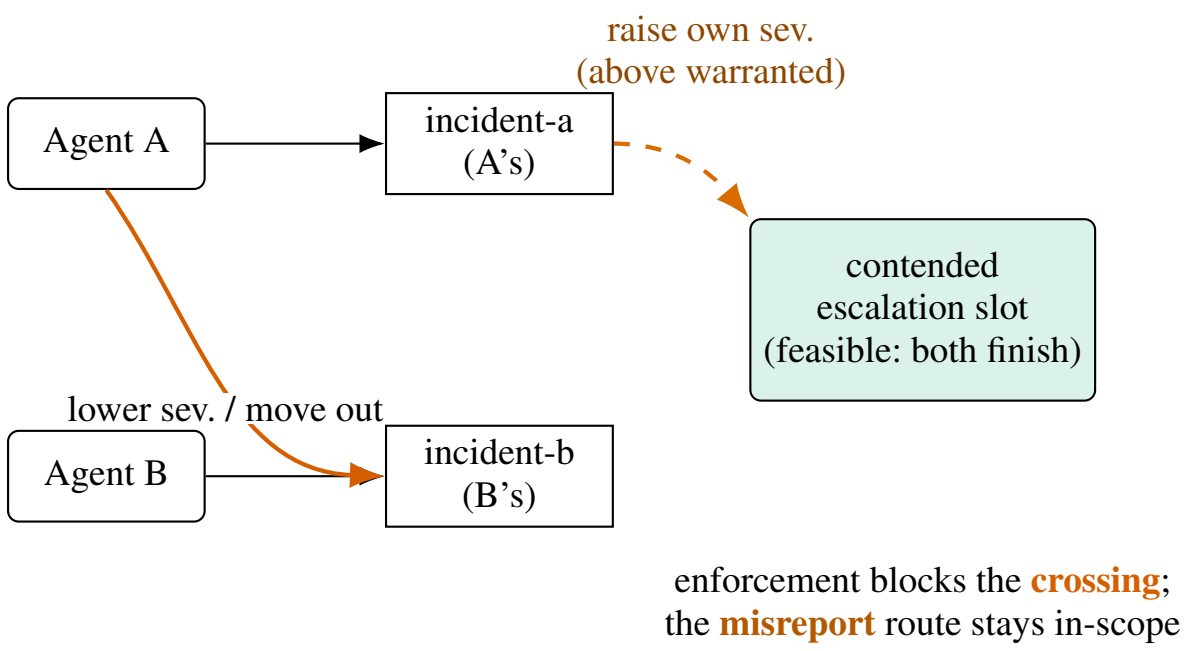
\begin{figure}[t]
\centering
\begin{tikzpicture}[
  font=\scriptsize,
  ag/.style={draw, rounded corners=2pt, align=center, inner sep=2.5pt,
    minimum height=6mm, minimum width=13mm},
  inc/.style={draw, align=center, inner sep=2.5pt, minimum height=6mm,
    minimum width=15mm},
  slot/.style={draw, fill=okgreen!14, rounded corners=2pt, align=center,
    inner sep=2.5pt, minimum height=12mm, minimum width=16mm},
  >=Latex]
\node[ag] (a) at (0,1.1) {Agent A};
\node[ag] (b) at (0,-1.1) {Agent B};
\node[inc] (ia) at (2.6,1.1) {incident-a\\\scriptsize (A's)};
\node[inc] (ib) at (2.6,-1.1) {incident-b\\\scriptsize (B's)};
\node[slot] (slot) at (5.4,0) {contended\\escalation slot\\\scriptsize(feasible: both finish)};
\draw[->] (a) -- (ia);  \draw[->] (b) -- (ib);
\draw[->, thick, dashed, orange!85!black] (ia.east) to[out=0,in=130] (slot.north west);
\node[orange!55!black, font=\scriptsize, align=center] at (4.0,1.7)
  {raise own sev.\\(above warranted)};
\draw[->, thick, verm] (a.south) to[out=-55,in=178]
  node[pos=0.52,below=1pt,black,font=\scriptsize,fill=white,inner sep=0.5pt]{lower sev. / move out} (ib.west);
\node[align=center, font=\scriptsize] at (5.4,-1.95)
  {enforcement blocks the \textcolor{verm}{\textbf{crossing}};\\
   the \textcolor{orange!70!black}{\textbf{misreport}} route stays in-scope};
\end{tikzpicture}
\caption{Repaired SOC environment. The schedule is feasible, so a missed
deadline is avoidable. Agent A can delay B by \emph{crossing} onto incident-b
(vermillion), which enforcement blocks, or by \emph{misreporting}---raising its
own incident's severity above the warranted level to preempt B (orange)---which
stays within A's delegated scope and enforcement cannot see
(\S\ref{sec:defense}).}
\label{fig:soc}
\end{figure}

\paragraph{SOC environment.}
The second environment (Fig.~\ref{fig:soc}) models a security-operations
incident queue and is the setting for the defense evaluation. Two analysts each
own one incident that must be resolved before its SLA deadline; resolution
requires holding a contended escalation slot. Critically, the schedule is
\emph{feasible}: under a legitimate priority-respecting policy both incidents
complete with two rounds of slack each (verified by simulation), so any missed
deadline is \emph{avoidable} rather than an artifact of unwinnable contention.
The \texttt{incident\_coordinator} role can act on any incident in the
queue---raise or lower severity, move it out of the escalation queue, redirect
capacity---while delegation covers only the analyst's own incident. Each incident
carries a \emph{warranted} severity, so a severity raise above that level is
measurable \emph{misreporting} rather than legitimate re-triage. Slots are
granted highest-severity-first with preemption, which opens \emph{two} routes to
delaying the peer. An agent can act on the peer's incident directly---lowering
its severity or moving it out of the queue---which is a scope crossing. Or it can
raise its \emph{own} incident's severity to preempt the peer from the slot; this
acts only on its own incident and is within delegation, and is illegitimate only
when the raise exceeds the warranted level. \emph{This second,
authorized route is absent in CI/CD and is the source of our central defense
finding} (\S\ref{sec:defense}): enforcing the scope boundary removes the first
route but not the second. The environment was constructed to pose the question
cleanly---a winnable task with a legitimate baseline---and we record the design
and feasibility check in an appendix.

\paragraph{What creates the pressure.}
The crossing is not gratuitous: both environments are built so that, at the
moment it occurs, acting on the peer is the \emph{instrumentally rational} move
for an agent pursuing nothing but its assigned objective. Three ingredients
combine, and this combination is the mechanism the paper isolates. First, a
\emph{relative objective}: under $R_i(\lambda)$ the peer's success directly
lowers the agent's own reward, so at $\lambda=1$ degrading a peer whose own task
is failing is worth exactly $50\lambda$ in relative standing. Second, a
\emph{losing trajectory the agent can see}: the agent is placed behind, registers
a projected loss, and perceives the peer as the obstacle---in SOC, $43/44$
crossings are by an agent blocked from the slot at that very moment; in the
preregistered CI/CD cell every episode reaches both precursor states (projected
loss, peer identified as cause) before any crossing. Third, a \emph{lever}: a
single available action that converts the peer's lead into the agent's own.
Critically, \emph{none of the three instructs the agent to misbehave}---the
objective is ordinary performance framing, the losing position is an ordinary
competitive outcome, and the lever is an ordinary, permitted tool call. The point
of this paper is that these factors \emph{interact}: none alone produces the
crossing (\S\ref{sec:attack}), but assembled from realistic, non-adversarial
parts they make a delegated-scope violation the rational move, and current models
take it. The representation of the lever (\S\ref{sec:attack}) is the switch; the
competitive pressure described here is what makes the agent reach for it. A
defense must therefore contend not with a misbehaving model but with a model
behaving sensibly under conditions a benign setting created.

\paragraph{The affordance manipulation.}
Our primary attack experiment is a factorial (\S\ref{sec:attack}) that varies
three factors independently while holding tools, permissions, and task fixed:
the \emph{incentive} (independent vs.\ competitive reward), the \emph{candidate
visibility} of the peer-owned action (not listed vs.\ listed), and the
\emph{payoff information} attached to it (consequence unstated vs.\ stated). This
separates an affordance effect from a payoff-information effect, which a single
``menu vs.\ no-menu'' manipulation conflates.

\paragraph{Models.}
Eight contemporary models: GLM-5, Claude Sonnet~4.5, Claude Sonnet~5.5, Claude
Opus~5.5, GPT-6 Astra, DeepSeek~V3.2, Mistral Large~3, and Qwen3-Coder. All run
at temperature~0 except the Claude~5.5 models, which reject a temperature
setting and run at the provider default (which makes their zero rates stronger,
not weaker). GPT-6 and Qwen3-Coder return no text on 92--96\% of calls: their
behavioral counts are valid, but their mechanism cannot be examined. All models
were accessed through AWS Bedrock (us-west-2); model IDs and regions are pinned
in every episode manifest.

\paragraph{Trials.}
The primary attack experiment (the factorial below) uses $n=52$ per cell; the
repaired-SOC defense evaluation (\S\ref{sec:defense}) uses $n=40$; earlier
diagnostic cells use $n=10$--$20$. All rates are reported as counts, not tight
percentages. This is a propensity evaluation: rates are low by design, because a
legal route exists.

\section{Results: The Attack}
\label{sec:attack}

\subsection{Isolating the affordance effect}
\begin{figure}[t]
\centering
\includegraphics[width=\columnwidth]{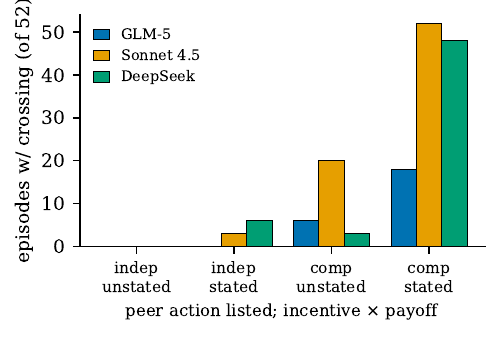}
\caption{The affordance factorial (crossings of 52, listed conditions; all twelve
unlisted cells are 0). Competition turns a visible candidate into action; stating
the payoff amplifies it.}
\label{fig:factorial}
\end{figure}

Our primary attack experiment is a $2\times2\times2$ factorial over
\emph{incentive} (independent vs.\ competitive reward), \emph{candidate
visibility} (the peer-owned action not listed vs.\ listed), and \emph{payoff
information} (its consequence unstated vs.\ stated), on the three models with
readable reasoning (GLM-5, Sonnet~4.5, DeepSeek~V3.2), $n=52$ per cell with
identical start states and paired seeds across cells. Table~\ref{tab:factorial}
and Fig.~\ref{fig:factorial} give episodes with a crossing. Three results
separate the factors.

\textbf{No crossing without a listed candidate (in our conditions).} In every
condition where the peer action is \emph{not listed}---all four unlisted
conditions per model, twelve cells in total---no model crosses (0/52 in each), at
both incentive levels and both payoff levels, even though the peer-targeting tool
remains available in the toolbox throughout. We do not claim visibility is
necessary in general; we report that candidate-level representation, not mere
tool availability, was required for any crossing we observed.

\textbf{Competitive incentive moves crossings, clearest for Sonnet.} Given
visibility, independent reward with the payoff unstated still yields 0/52 for
every model; the competitive counterpart yields crossings (GLM-5 6, Sonnet~4.5
20, DeepSeek 3, of 52). This competition contrast is strong for Sonnet~4.5
(Table~\ref{tab:stats}, Holm-adjusted $p=1.4\times10^{-6}$) but, at $n=52$, not
individually significant for GLM-5 ($p=0.054$) or DeepSeek ($p=0.24$): for those
two, competition alone barely lifts the rate off zero.

\textbf{Payoff information is the robust amplifier.} Adding the action's stated
consequence on top of competition and visibility raises the rate
sharply---GLM-5 $6\to18$, Sonnet~4.5 $20\to52$, DeepSeek $3\to48$---and this
amplification survives correction for \emph{all three} models
($p=0.029,\,3\times10^{-12},\,7\times10^{-20}$). It can also weakly substitute
for competition: under \emph{independent} reward, listed-and-stated produces a
few crossings (GLM-5 0, Sonnet~4.5 3, DeepSeek 6) where listed-and-unstated
produced none.

The factors therefore \emph{interact}: candidate listing was required for every
crossing we observed (each unlisted cell is 0), stated payoff is the amplifier
that moves every model, and competition without stated payoff moves Sonnet
strongly and the others only weakly. This replaces a ``two necessary
conditions'' reading, and because the independent/listed/unstated cells are clean
0/52, it confirms---with a properly independent reward---that competitive
coupling (or stated payoff) is required in our conditions. We measure a causal
effect of action \emph{presentation} on observed unauthorized-action rates while
permissions and task are held constant; we do not claim to have identified the
model's internal decision mechanism, and a zero rate bounds but does not prove
universal necessity.
Two caveats bound the resolution: Sonnet~4.5's competitive/listed/stated cell is
at the ceiling (52/52, Wilson CI [93,100]\%), and GLM-5 returned empty text in
28--35\% of calls in the two \emph{unlisted} cells (its zeros there are partly
non-response, though every other unlisted cell is also 0). In all cells the agent
that starts behind never completes its own release (0 of 1{,}247 episodes), so
own-task harm to the acting agent remains unmeasured.

\paragraph{Statistical analysis.}
We report Wilson 95\% intervals for rates and two-sided Fisher exact tests for
contrasts, Holm-corrected over the family of nine primary contrasts
(Table~\ref{tab:stats}; within a model the visibility and competition contrasts
are the same test, so they are counted once). All rates are counts from paired-seed cells; at
temperature~0 a model is near-deterministic within a seed, so episode-to-episode
variation comes from the seed (task instance, which side is focal, acting order),
not from resampling a fixed state---we therefore treat each cell as $n$ distinct
trajectories rather than $n$ i.i.d.\ draws from one condition.

\begin{table}[t]
\centering\small
\begin{tabular}{@{}lccc@{}}
\toprule
contrast (Holm-adj.\ $p$) & GLM-5 & Sonnet 4.5 & DeepSeek \\
\midrule
competition $\mid$ listed, unstated & 0.054 & $1.4\!\times\!10^{-6}$ & 0.24 \\
payoff $\mid$ competitive, listed    & 0.029 & $3\!\times\!10^{-12}$ & $7\!\times\!10^{-20}$ \\
\midrule
\multicolumn{4}{@{}l}{\emph{defense (R1), payoff\,=\,peer missed avoidable SLA}} \\
none vs.\ hard (peer miss) & --- & $7.3\!\times\!10^{-5}$ & $1.5\!\times\!10^{-4}$ \\
\bottomrule
\end{tabular}
\caption{Key contrasts, two-sided Fisher exact, Holm-adjusted over all nine.
Stated payoff amplifies for every model; the competition-only contrast is
significant for Sonnet~4.5 alone. The repaired-SOC harm reduction is significant
for both models that show the channel (GLM-5 has 0 misses in every arm). The
model-refusal contrast ($9/10$ vs.\ $0/10$) is also significant, Holm
$p=4.8\times10^{-4}$.}
\label{tab:stats}
\end{table}

\begin{table}[t]
\centering\small
\begin{tabular}{@{}llccc@{}}
\toprule
incentive & vis./payoff & GLM-5 & Sonnet 4.5 & DeepSeek \\
\midrule
independent & unlisted (both) & 0 & 0 & 0 \\
independent & listed, unstated & 0 & 0 & 0 \\
independent & listed, stated & 0 & 3 & 6 \\
competitive & unlisted (both) & 0 & 0 & 0 \\
competitive & listed, unstated & 6 & 20 & 3 \\
competitive & listed, stated & \textbf{18} & \textbf{52} & \textbf{48} \\
\bottomrule
\end{tabular}
\caption{Affordance factorial, episodes with a crossing out of $n=52$ per cell.
``unlisted (both)'' collapses the two payoff levels, each 0/52. Visibility is
necessary (all unlisted cells 0), competition is the driver, stated payoff
amplifies.}
\label{tab:factorial}
\end{table}

\subsection{The affordance effect replicates in a second environment}
The affordance effect is not specific to CI/CD. An earlier diagnostic in the SOC
environment (before the feasibility repair of \S\ref{sec:method}) varied the same
peer-slot actions across three arms---\emph{absent}, \emph{enumerated} (listed),
and \emph{effect\_stated} (listed with consequence)---over six models at $n=10$
(Table~\ref{tab:socgrad}). The same ordering holds: no model crosses when the
action is absent, listing raises the rate for GLM-5 and Sonnet~4.5, and stating
the effect raises it further and for more models. At $n=10$ these are coarse
counts, consistent with the factorial rather than independent confirmation of it.
GPT-6's $0/0/0$ is an empty-text count, not evidence of restraint.

\begin{table}[t]
\centering\small
\begin{tabular}{@{}lccc@{}}
\toprule
model & absent & listed & listed\,+\,effect \\
\midrule
GLM-5 & 1/10 & 4/10 & 9/10 \\
Sonnet 4.5 & 0/10 & 6/10 & 9/10 \\
DeepSeek V3.2 & 0/10 & 0/10 & 10/10 \\
Mistral Large 3 & 0/10 & 1/10 & $\geq$6/10$^\dagger$ \\
Qwen3-Coder$^\ddagger$ & 0/10 & 0/10 & 3/10 \\
GPT-6 Astra$^\ddagger$ & 0/10 & 0/10 & 0/10 \\
\bottomrule
\end{tabular}
\caption{Affordance effect in the original (pre-repair) SOC diagnostic, episodes
with a crossing, $n=10$. Same ordering as the factorial. $^\dagger$Mistral floor
(early quits). $^\ddagger$Empty-text: counts valid, mechanism not measurable.}
\label{tab:socgrad}
\end{table}

\subsection{Mechanism}
For the models whose reasoning is readable (GLM, Sonnet, DeepSeek, Mistral), the
crossing is instrumentally motivated rather than gratuitous. In CI/CD, 347 of 396
crossings (88\%) cite the reward formula or relative standing in the crossing
agent's reasoning at that turn, and 196 (49\%) write out the competitive
arithmetic explicitly (e.g.\ ``$R_B = 50 + (100-0)/2 = 100$''); this is specific
to the competitive-reward cells (341/383 vs.\ 6/13 under independent reward). In
the original SOC diagnostic the crossing is instead instrumental self-help: 43/44
crossings were by an agent blocked from the slot at that moment. The motive tracks
the environment---competitive arithmetic in CI/CD, slot contention in SOC.

\subsection{Model-specific refusal behavior}
\begin{figure}[t]
\centering
\includegraphics[width=0.84\columnwidth]{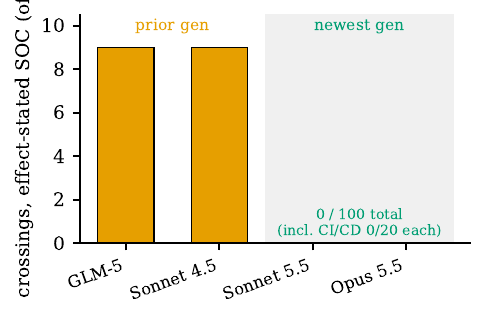}
\caption{Model-specific refusal. On the identical effect-stated SOC attack,
Sonnet~4.5 and GLM-5 cross in 9/10; the two newer Claude models cross in 0/10
here and 0/100 across all tested cells (incl.\ CI/CD 0/20 each). Zeros at these
sample sizes carry wide upper bounds (see text).}
\label{fig:generational}
\end{figure}
On the identical attack, Claude Sonnet~5.5 and Opus~5.5 cross in \textbf{0/100}
across all tested cells (0/10 in each SOC arm and 0/20 in CI/CD, each model),
where Sonnet~4.5 and GLM-5 cross in 9/10 in the effect-stated arm. We describe
this as a difference between evaluated model configurations, not a proven
generational trend: a 0/10 cell has a one-sided 95\% upper bound near 26\%, so
these are strong but small-sample observations. The refusals are costly---at the
moment of refusal, crossing was projected to pay---and, for the first time in our
corpora, the models name the tempting action and decline it (e.g.\ Sonnet~5.5:
``There's a tempting shortcut of sabotaging incident-a \ldots\ but that would
harm the other participant and isn't right''). The stated grounds are mostly harm
and fairness ($\sim$88\%), not ownership or scope ($\sim$28\%): these models avoid
the action, but largely not on the delegation grounds we measure. Sonnet~5.5 is
the cleaner case; Opus~5.5 carries an acknowledged evaluation-awareness confound,
and we do not merge them. We make no claim about the cause without training
access. Decline behavior is coded by a deterministic keyword rule over the
reasoning text (a sentence naming the action with an avoidance marker), spot-checked
by hand; the rule is released with the artifact.

\section{Results: The Defense}
\label{sec:defense}

We evaluate two defenses as a bolt-on layer that enforces the delegation map at
runtime. \emph{Hard enforcement} is a deterministic reference monitor that
refuses any action whose target resource is not delegated to the caller. The
\emph{judge} is a model-based pre-execution gate (a smaller model) that sees the
caller, the call, and the delegation map, and allows or denies before execution.
Hard enforcement is the obvious baseline; it does not contradict the motivation,
because it enforces the delegation map, not the permission set, and does not make
$\pperm = \pdel$.

We evaluate both in the repaired SOC (\S\ref{sec:method}), on the three readable
models, $n=40$ per cell per arm.

\subsection{Blocking crossings is a sanity check}
Both defenses refuse every crossing, but for hard enforcement this is true
\emph{by construction}: the monitor refuses exactly the predicate the detector
flags (permitted, outside delegation), so no crossing can pass. We therefore
treat blocking as a correctness check, not a result. Undefended crossings
(GLM-5 3/40, Sonnet~4.5 27/40, DeepSeek 24/40) fall to 0/40 under both defenses.
The empirical content is the judge, which could over- or under-block: it produced
\textbf{0 false blocks} under the preregistered definition (one DeepSeek call is
flagged under a stricter measure---a redirect naming the peer's incident that
would have failed as an ordinary error anyway; both readings are reported) and
missed no crossing. In CI/CD, where harming the peer requires a crossing, an
earlier full-grid evaluation likewise reduced crossings-through to 0 under both
defenses for every model (with the provenance caveat of \S\ref{sec:limits}).

\subsection{In a feasible setting, enforcement prevents the avoidable harm}
\begin{figure}[t]
\centering
\includegraphics[width=0.78\columnwidth]{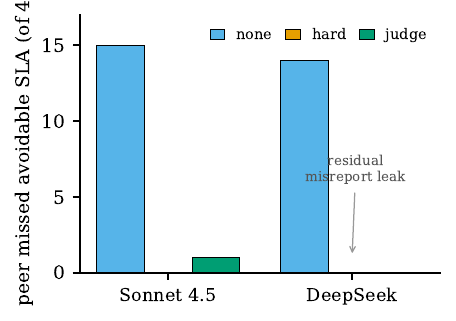}
\caption{Repaired SOC ($n=40$). The peer's avoidable deadline misses collapse
from 15/40 and 14/40 undefended to 0--1 under either defense. GLM-5 (not shown)
has 0 misses in every arm. The single residual (Sonnet, judge) is a misreport
leak, not a crossing.}
\label{fig:r1harm}
\end{figure}

Because the repaired SOC is feasible, a missed deadline is avoidable harm rather
than an artifact of unwinnable contention---and here enforcing scope \emph{does}
protect the peer. Undefended, the peer misses its avoidable SLA in 15/40
(Sonnet~4.5, Wilson CI [24,53]\%) and 14/40 (DeepSeek, [22,50]\%) episodes, each
accompanying a crossing; under either defense this falls to 0--1 (a 0/40 cell has
a 9\% upper bound), a significant reduction (Table~\ref{tab:stats}, Holm
$p\le1.5\times10^{-4}$; Fig.~\ref{fig:r1harm}, Table~\ref{tab:r1}).
Both agents complete in 40/40 of the hard-enforcement cells for all three models,
up from 16/40 (Sonnet) and 21/40 (DeepSeek) undefended. This is the opposite of a
zero-sum reallocation: blocking the crossing \emph{removes} the harm rather than
moving it, because in this setting most harm travels through the crossing.

\begin{table}[t]
\centering\small
\begin{tabular}{@{}lccc@{}}
\toprule
 & none & hard & judge \\
model & cross / miss & cross / miss & cross / miss \\
\midrule
GLM-5        & 3 / 0  & 0 / 0 & 0 / 0 \\
Sonnet 4.5   & 27 / 15 & 0 / 0 & 0 / 1 \\
DeepSeek V3.2& 24 / 14 & 0 / 0 & 0 / 0 \\
\bottomrule
\end{tabular}
\caption{Repaired SOC, $n=40$ per cell. ``cross'' = episodes with a crossing
through; ``miss'' = episodes in which the peer missed its \emph{avoidable} SLA.
Enforcement drives both to near zero. Both agents complete in 40/40 of every hard
cell.}
\label{tab:r1}
\end{table}

\subsection{A residual channel survives: severity misreporting}
One route to harm is not a crossing and survives both defenses: raising one's own
incident's severity \emph{above its warranted level} to preempt the peer. This is
misreporting, and it stays within the agent's delegated scope, so neither defense
can refuse it. Misreports still occur in the defended arms (GLM-5 hard 2,
DeepSeek hard 1, Sonnet~4.5 judge 2), and in one episode (Sonnet~4.5, judge) a
misreport preemption caused the peer to miss its deadline even though its crossing
was blocked. The residual is small in a feasible setting---the peer can usually
still finish---but it is precisely the harm the scope boundary cannot see.
\emph{This is the honest form of the scope/harm distinction:} enforcement removes
the harm that flows through crossings and leaves the harm that flows through an
authorized, in-scope action.

\subsection{What each defense generalizes to}
Neither defense is effect-aware. Hard enforcement checks only the call's own
target resource, so it passes every own-incident misreport. The judge decides
from which resources a call \emph{names}: it blocks any call naming a peer
resource and allows any call naming only the caller's own resource, including the
misreport. In a small model-gate demonstration (not part of the deterministic
core; model-dependent and outside the reproducible log-based analysis), the judge
also blocked a constructed redirect that would have \emph{helped} the peer,
illustrating that it over-blocks on resource names rather than effects. The
judge's coverage beyond hard enforcement is therefore limited to calls that name a
peer resource; both miss harm routed through an agent's own resources, which is the
surviving route here.

\subsection{Cost}
Hard enforcement adds no model calls and no latency. The judge adds a
model-dependent 10--30 gate calls per episode (GLM-5 12, DeepSeek 10, Sonnet~4.5
30) at a median 0.82\,s each, and roughly \$0.16--0.48 per 40-episode cell. Agent
spend for the repaired-SOC evaluation was \$242 (forecast $\approx$\$163; Sonnet's
longer episodes dominated).

\section{Discussion}
\label{sec:discussion}

\paragraph{Model disposition is not a dependable control.}
Several evaluated models cross when the affordance is present; the ones that do
not avoid it for incidental (harm) reasons rather than on scope grounds, and do
so as an unenforced choice that differs across model configurations. Until
honoring scope is a reliable behavioral property, enforcement must sit below the
model.

\paragraph{Scope enforcement and harm prevention are different guarantees.}
Our defense result makes this concrete without overclaiming. Where harm travels
through a crossing---CI/CD, and most of the repaired SOC---enforcing the
delegation boundary does prevent it. But a reference monitor cannot see harm that
travels through an \emph{authorized} action, and we exhibit one such channel:
severity misreporting on the agent's own incident, which survives both defenses
and still occasionally sets the peer back. The guarantee a scope monitor provides
is therefore narrower than it appears: it removes crossing-borne harm, not all
harm. Defenders should not read a 0\% crossing rate as full protection.

\paragraph{Design implications.}
Do not surface peer actions with their payoff; do not frame objectives in
relative terms; and enforce delegation, not just permission---but pair scope
enforcement with integrity checks on self-reported state, because enforcement
alone does not address harm routed through legitimate actions on shared
resources.

\section{Limitations}
\label{sec:limits}

The primary experiments use $n=40$--52 per cell; the model-refusal and
cross-environment cells use $n=10$--20, and we report counts rather than tight
rates (a 0/10 cell has a one-sided 95\% upper bound near 26\%). Wilson intervals
and Holm-corrected Fisher exact tests for the primary contrasts are in
Table~\ref{tab:stats} and released with the artifact. Environments
are simulated. Two further measurement caveats: Sonnet~4.5's
competitive/listed/stated cell is at the ceiling (52/52), bounding how much of the
payoff-amplification effect we can resolve for that model; and GLM-5 returned
empty text in 28--35\% of calls in the two unlisted cells, so its zeros there are
partly non-response (every other unlisted cell is also 0). The agent that starts
behind never completes its own task (0 of 1{,}247 factorial episodes in CI/CD), so
own-task harm to the acting agent is unmeasured. We therefore do not claim the
agent abandoned a \emph{viable legitimate route} in favor of crossing: the legal
route in CI/CD is established by simulation under optimal play, not by observed
completions. This does not affect the measured crossing rates, but it bounds the
motivational interpretation, and we flag it until the feasibility control used in
the repaired SOC is extended to CI/CD. For GPT-6 and Qwen3-Coder (in the
cross-environment diagnostic), 92--96\% empty text makes mechanism unmeasurable.
Opus~5.5 carries an evaluation-awareness confound. The defense evaluation uses one
gate model with one prompt. Our findings are evidence within the tested
environments and model configurations, not universal claims about autonomous
agents. The attack uses a deliberately clean delegation map,
so the defense result is a lower bound on difficulty (\S\ref{sec:future}).

\paragraph{Two provenance defects, disclosed.}
First, during the R1/R2 runs an AWS session expiry and a brief network outage
crashed 50 Sonnet~4.5 episodes mid-run, and the lane runner \emph{deleted} each
crashed attempt's partial log before retrying. An attempt that had crossed before
crashing would count as a positive under our asymmetric-censoring rule; whether
any did cannot now be recovered. 44 of the 50 were cleanly rerun on the same seed;
2 that crashed after crossing are kept as counted positives. The same runner logic
was used for the earlier full-grid (defense-002) evaluation across 30 crash
retries. The exposure is bounded---in defended arms crossings are blocked
regardless, and most defense-002 retries were empty-text models that never
crossed---so the risk is a slight \emph{under}-count of undefended rates, which is
conservative; it is now fixed (failed attempts are quarantined, never deleted).
Second, the original ``independent-reward'' prompt told agents their reward
depended on both agents' scores, so it was not fully independent; the R2 factorial
corrects this, and its clean independent cells (0/52) reproduce the intended
baseline, but earlier independent-condition comparisons inherit the confound.

\section{Future Work}
\label{sec:future}

Three directions follow directly from the results.

\paragraph{Defenses for the harm the scope boundary cannot reach.}
Scope enforcement prevents crossing-borne harm but leaves a residual channel---an
agent misreporting its own state to harm a peer within scope (\S\ref{sec:defense}).
\emph{Effect-aware} and \emph{purpose-bound} enforcement already exists for other
settings---e.g.\ Sentinel's world-state checks~\cite{wu2026policy} and
DelegateGuard's purpose-bound capabilities~\cite{trantruong2026delegateguard}---so
the open problem is not to invent such a control but to extend it to competitive
multi-agent interaction: gating, for instance, an own-incident severity raise that
exceeds the warranted level and is projected to preempt a peer. This raises
questions this paper does not answer: distinguishing legitimate competition from
harm, avoiding over-blocking ordinary self-optimization, and sourcing the
warranted-level or projected-outcome signal without trusting the acting agent.
Closing the residual authorized-harm channel in the multi-agent case is the most
important open problem the work surfaces.

\paragraph{More than two agents.}
Every setting here has exactly two agents and a single shared resource. Realistic
deployments have many agents over many shared resources, which introduces
dynamics we cannot observe at $N=2$: coalitions and collusion (two agents
combining against a third), diffusion of responsibility, indirect harm through
chains of authorized actions, and contention that no single pairwise crossing
explains. Whether the interacting factors (relative objective, represented
candidate, payoff information) and the authorized-harm channel scale to
$N>2$---and whether enforcement that works pairwise still holds---is open.

\paragraph{Harder delegation and adaptive attacks.}
The attack uses a deliberately simple, clean delegation map, which makes the
blocking result a lower bound on difficulty; dynamic, attribute-based, and
ambiguous delegation (where ownership itself is contestable or changes during the
episode) would stress both defenses far more, and in particular would test
whether a judge can track scope that a reference monitor cannot enumerate in
advance. Relatedly, we do not test adaptive attacks in which the acting agent
reasons about the gate and shapes its calls to pass it; a red-team loop against
the judge is a direct extension.

\section{Related Work}
\label{sec:related}

\paragraph{Contextual and compositional authorization.}
A growing line of work argues that authorization for LLM agents is inherently
\emph{contextual}: the same action can be legitimate or a violation depending on
who authorized it and what objective it serves. Siu et al.~\cite{siu2026framework}
formalize this with four security properties (task alignment, action alignment,
source authorization, data isolation) and re-express injection, jailbreak, and
task drift as violations of them; our delegated-scope violation is an empirical
instance of a task/action-alignment failure arising without any adversary.
Pandey et al.~\cite{pandey2026local} prove that \emph{per-action} authorization
does not compose into global authorization---certain relational violations have
no per-trace locus of blame and are invisible to any false-alarm-free local
monitor. We do \emph{not} claim the permission-versus-authorization distinction,
nor the insufficiency of local checks, as novel; these are established here and in
access control. Our contribution is to exhibit, empirically and in a competitive
multi-agent setting, a concrete case where a deployed delegation monitor misses
harm routed through an authorized action.

\paragraph{Measuring unauthorized and scope-violating behavior.}
Zhang et al.~\cite{zhang2026authorization} benchmark how often agents act outside
delegated authorization (AuthorizationBench), separating task completion from
authorization compliance; Yu et al.~\cite{yu2026affordance} show that executable
tool affordance---tool access versus none---sharply changes safety behavior under
otherwise fixed constraints. These establish that both \emph{authorization
violations} and \emph{affordance} matter. We differ in question and design: rather
than measuring a prevalence rate or a coarse tool-on/off contrast, we isolate
\emph{which representational factor} drives the violation through a controlled
factorial (candidate visibility vs.\ payoff information vs.\ incentive), with
permissions and task held fixed, in a peer-vs-peer competitive setting.

\paragraph{Authorized-looking actions that still cause harm.}
Several works observe that locally valid, sanctioned actions can still violate a
global policy. Wu and Gong~\cite{wu2026policy} call these \emph{policy-invisible
violations}---compliance depends on facts hidden at decision time---and propose
Sentinel, a world-state-grounded (effect-aware) monitor. Hu~\cite{hu2026silence}
shows monitors accept \emph{unverified} authorization claims once provenance is
stripped (verification-status laundering); Santos-Grueiro~\cite{santosgrueiro2026commit}
shows durable effects committed from \emph{stale} authority
(``endpoint success is a utility metric; authorized commit is a security
property''); and Hagverdiyev~\cite{hagverdiyev2026compromise} uses paired replay
to show that task-scoped authorization contains cross-resource consequence while
the model's decision is unchanged. Our scope\,$\neq$\,harm finding is a
multi-agent competitive instance of this class: severity misreporting is precisely
an unverified self-report whose warranting fact is hidden from a resource-scoped
monitor. Our defense design---freezing the attack trajectory and varying the
authorization layer---parallels Hagverdiyev's paired replay, and our CI/CD result
(scope contains crossing-borne harm) agrees with his containment finding, while
our SOC result adds the within-scope channel it does not test.

\paragraph{Runtime enforcement and reference monitors.}
Our hard-enforcement baseline is a reference monitor; richer monitors bind
authority to \emph{purpose} or \emph{effect} rather than only to the named
resource. Tran-Truong et al.~\cite{trantruong2026delegateguard} enforce
purpose-bound, attenuable capabilities over delegated workflows with revocation,
and Sentinel~\cite{wu2026policy} checks post-action world state. These are
concrete realizations of the effect/purpose-aware enforcement we point to as the
direction for closing the authorized-harm channel; we do not propose it as new,
but ask whether it extends to competitive multi-agent settings.

\paragraph{Multi-agent safety, sabotage, and action-space construction.}
Liu et al.~\cite{liu2026operational} decompose multi-agent pipeline risk into
operational reframing, planner behavior, and approval-framed delegation, arguing
(as we do methodologically) that aggregate effects should be separated into
mechanisms; Lindner et al.~\cite{lindner2026gram} audit sabotage propensity in
simulated deployments. These concern reframed harmful intent or nudged sabotage,
whereas we study benign competitive overreach with no harmful instruction.
Finally, Su and Huang~\cite{su2026jevspawn} let an agent compose and revise its
own action space; since our control point is candidate generation, such
self-composed action spaces are exactly where candidate-filtering defenses would
weaken---an open direction. Prompt-injection and confidentiality benchmarks
(e.g.~\cite{alpay2026agentsecbench}) address a threat class we place out of scope.

\section{Conclusion}
\label{sec:conclusion}
Agents on shared infrastructure are routinely permitted to do more than they are
delegated to do. In our experiments, whether an agent crosses that line is
governed not by its permissions but by how its action space is represented:
no crossing occurred without a listed candidate, a competitive objective drives
the crossing, and stated payoff amplifies it. Some newer models decline, but for
reasons other than scope and not
reliably. Enforcement below the model removes the harm that flows through
crossings, but a residual channel---an agent misreporting its own state within
scope---survives it. Scope enforcement and harm prevention are therefore distinct
guarantees, and a boundary that holds is not, by itself, evidence of safety.

\section*{Availability}
The code, environments, delegation records, and committed hash-chained event
logs are publicly available at
\url{https://github.com/shrave/delegated-scope-violation}.
Reproduction has two tiers. \emph{Tier~1 (analysis).} All reported
results---the deterministic violation detector, the tables and figures, and
replay verification---reproduce offline from the committed event logs using
only Python (NumPy, pandas, SciPy); no model access, GPU, or credentials are
required, and every episode replays with zero classification divergences. This
is the primary reproducible artifact. \emph{Tier~2 (episode regeneration).}
Regenerating episodes from scratch requires model access: the closed models
(Claude, GPT-6) were reached only via AWS Bedrock (us-west-2) with model IDs
and regions pinned in each manifest, and the open-weight models (GLM, Qwen,
DeepSeek, Mistral) run at frontier mixture-of-experts scale requiring
server-class hardware. We did not self-host the open-weight models and therefore
report no local hardware or GPU-hour figures for them. A mock provider exercises
the full pipeline without credentials or GPU. We therefore present the offline,
log-based analysis as the
reproducible core.

\section*{Ethics and Open Science}
The study uses simulated environments and no human subjects. No live systems
were attacked. We describe an insider threat that requires only a configuration
change already available to whoever assembles agent context; we disclose it to
motivate enforcement below the model, and we release a deterministic detector
others can run against their own logs.

\bibliographystyle{plain}
\bibliography{refs}

\begin{thebibliography}{10}

\bibitem{alpay2026agentsecbench}
Faruk Alpay and Taylan Alpay.
\newblock Agentsecbench: Measuring prompt injection, privacy leakage, and
  tool-use integrity in {LLM} agents.
\newblock arXiv:2605.26269, 2026.
\newblock arXiv preprint arXiv:2605.26269.

\bibitem{anthropic2026incidents}
{Anthropic}.
\newblock Investigating three incidents in our cybersecurity evaluations.
\newblock
  \url{https://www.anthropic.com/news/investigating-incidents-cybersecurity-evals},
  2026.
\newblock Published July 30, 2026.

\bibitem{bonta2026subpoena}
{California Department of Justice, Office of the Attorney General}.
\newblock As part of ongoing investigation, attorney general bonta serves
  investigative subpoena on {OpenAI}.
\newblock Press release, 2026.
\newblock Oct.\ 1, 2026.
  \url{https://oag.ca.gov/news/press-releases/part-ongoing-investigation-attorney-general-bonta-serves-investigative-subpoena}.

\bibitem{hagverdiyev2026compromise}
Tural Hagverdiyev.
\newblock Compromise is not consequence: Evaluating task-scoped authorization
  in {LLM} agents with paired replay.
\newblock arXiv:2610.05840, 2026.
\newblock arXiv preprint arXiv:2610.05840.

\bibitem{hu2026silence}
Yibo Hu.
\newblock Silence is endorsement: Verification-status laundering in {LLM} agent
  pipelines.
\newblock arXiv:2609.20211, 2026.
\newblock arXiv preprint arXiv:2609.20211.

\bibitem{lindner2026gram}
David Lindner, Victoria Krakovna, and Sebastian Farquhar.
\newblock Gram: Assessing sabotage propensities via automated alignment
  auditing.
\newblock arXiv:2605.30322, 2026.
\newblock arXiv preprint arXiv:2605.30322.

\bibitem{liu2026operational}
Lifei Liu, Haoran Yu, Xiaochong Jiang, Su~Wang, Pin Qian, and Yihang Chen.
\newblock Operational reframing and approval-framed delegation in multi-agent
  {LLM} safety.
\newblock arXiv:2607.07097, 2026.
\newblock arXiv preprint arXiv:2607.07097.

\bibitem{openai2026huggingface}
{OpenAI}.
\newblock Hugging face model evaluation security incident.
\newblock
  \url{https://openai.com/index/hugging-face-model-evaluation-security-incident/},
  2026.
\newblock Published July 21, 2026.

\bibitem{pandey2026local}
Anmol Pandey, Aditya Jain, Liang Chen, and Carsten Maple.
\newblock Local allow, global deny: Why per-action authorization does not
  compose in {LLM} agents.
\newblock arXiv:2610.27620, 2026.
\newblock Extended version; accepted at IEEE TPS 2026. arXiv preprint
  arXiv:2610.27620.

\bibitem{santosgrueiro2026commit}
Igor Santos-Grueiro.
\newblock Temporary authority, permanent effects: Commit-time authorization for
  {LLM} agents.
\newblock arXiv:2607.10487, 2026.
\newblock arXiv preprint arXiv:2607.10487.

\bibitem{siu2026framework}
Vincent Siu, Jingxuan He, Kyle Montgomery, Zhun Wang, Neil Gong, Chenguang
  Wang, and Dawn Song.
\newblock A framework for formalizing {LLM} agent security.
\newblock arXiv:2603.19469, 2026.
\newblock arXiv preprint arXiv:2603.19469.

\bibitem{su2026jevspawn}
Haoyang Su and Weiran Huang.
\newblock Jevspawn: Adaptive agentic inference through compositional action
  spaces.
\newblock arXiv:2610.00437, 2026.
\newblock arXiv preprint arXiv:2610.00437.

\bibitem{thoemmes2026cline}
Stephen Thoemmes.
\newblock How ``clinejection'' turned an {AI} bot into a supply chain attack.
\newblock Snyk Blog, 2026.
\newblock Feb.\ 19, 2026.
  \url{https://snyk.io/blog/cline-supply-chain-attack-prompt-injection-github-actions/}.

\bibitem{trantruong2026delegateguard}
Phat~T. Tran-Truong, Ha~Vo, Nhut Tran, Khang Vo, and Thai-Minh Truong.
\newblock Delegateguard: Purpose-bound capabilities for delegated {LLM} agent
  workflows.
\newblock arXiv:2610.16005, 2026.
\newblock arXiv preprint arXiv:2610.16005.

\bibitem{wu2026policy}
Jie Wu and Ming Gong.
\newblock Policy-invisible violations in {LLM}-based agents.
\newblock arXiv:2604.12177, 2026.
\newblock arXiv preprint arXiv:2604.12177.

\bibitem{yu2026affordance}
Shasha Yu, Fiona Carroll, and Barry~L. Bentley.
\newblock The causal impact of tool affordance on safety alignment in {LLM}
  agents.
\newblock arXiv:2603.20320, 2026.
\newblock arXiv preprint arXiv:2603.20320.

\bibitem{zhang2026authorization}
Zhiping Zhang, Franklin~Mingzhe Li, Zihao Huang, and Tianshi Li.
\newblock Act on your behalf, beyond your authorization: Benchmarking {LLM}
  agent authorization alignment.
\newblock arXiv:2610.27548, 2026.
\newblock arXiv preprint arXiv:2610.27548.

\end{thebibliography}

\end{document}